\documentclass[12pt]{article}
\usepackage{amssymb}
\usepackage{graphicx}
\usepackage{array}
\usepackage{color}
\usepackage{bigstrut}
\usepackage{amsmath}
\usepackage{bm}
\usepackage{enumerate}
\usepackage{epsfig}
\begin{document}


\title { Influence of Hydrodynamic Fluctuations on the 
Phase Transition in Models E and F of Critical Dynamics.}

\author{M. Dan\v{c}o$^1$,  M. Hnatich$^{1,2}$,  M. V. Komarova $^{3}$, D. M. Krasnov$^3$,\\
  T. Lu\v{c}ivjansk\'y$^{1,2}$,  L. Mi\v{z}i\v{s}in$^2$,
M. Yu. Nalimov$^3$ 
\\
 $^{1}$ Institute of Experimental
Physics, Slovak Academy of Sciences,\\ Watsonova 47, 040 01
Ko\v{s}ice, Slovakia,\\
$^{2}$ Faculty of Sciences, P.J. \v{S}afarik
University,\\ Moyzesova 16, 040 01 Ko\v{s}ice, Slovakia,\\
$^3$ Department of Theoretical Physics, St. Petersburg University,\\Ulyanovskaya 1, St. Petersburg, Petrodvorets, 198504 Russia.}

\maketitle

\begin{abstract}
We use the renormalization group method to study model E of critical dynamics 
 in the presence of velocity fluctuations arising in accordance with the stochastic
  Navier-Stokes equation. Using Martin-Siggia-Rose theorem, we obtain a field-theoretical
  model that allows a perturbative renormalization group analysis.
By direct power counting and an analysis of ultraviolet divergences, we show
that the model is multiplicatively renormalizable, and we use a two-parameter expansion in
 $\epsilon$ and $\delta$ to calculate renormalization constants. Here, $\varepsilon$ is
 a deviation from the critical
 dimension four, and $\delta$ is a deviation from the Kolmogorov regime.
 We present the results of the one-loop approximation and part of the fixed-point
 structure. We briefly discuss the possible effect of velocity fluctuations on the 
 large-scale behavior of the model.
\end{abstract}

\section {Introduction}

Bose condensation is an  important physical phenomenon observed nowadays 
 not only in the superfluidity of liquid helium but also in the condensation of inert 
 gases \cite{Abrikosov}.
According to \cite{Hohenberg}, the critical dynamics near such a phase transition
 can be  described using model F. This model was analyzed in 
   \cite{Dominicis} using the renormalization group (RG) approach. 
It was shown that in the critical region,  model  F is equivalent to  model E
(according  to the standard terminology introduced in \cite{Hohenberg} ) 


Both dynamical models E and F of critical dynamics are free from 
hydrodynamic modes because velocity field turns out to be infrared (IR)  irrelevant in the critical range. 
Therefore, the critical exponent (e.g. for the viscosity) is still unknown, although the viscosity 
vanishes  during the considered phase transition and manifests  the features of order parameter. 
Moreover,  the problem of the influence of turbulence on the phase transition into the 
superfluid state remains unsolved. 

A stochastic equation for the critical dynamics of a Bose system in the presence of a random velocity 
field was proposed in \cite{Komarova}. Such a modification of model E leads
to some deviations from the standard field-theoretical approach, and we adopt it.
We here continue the investigation begun in \cite{Komarova}.  Our aim is to study 
different scaling regimes of the proposed model. 

This paper is structured as follows. We begin by  analyzing the field theory formulation
of the model and its renormalization (see Sec. 2).  In Sec. 3, we present
some interesting details of the one-loop calculation   
and give relations between renormalization constants.  In Sec. 4, we analyze the fixed
points and their regions of IR stability and give the results of 
 the one-loop calculations of the RG functions. In Sec. 5, we present brief conclusions.
\section {Field-theoretic formulation of the model}
The stochastic equations of Bose-like systems can be described in the vicinity 
of their critical points \cite{Komarova} by the equation
\begin{align}
  \partial_t\psi + \partial _i(v_i\psi) &  =  
  \lambda  (1+ib)[\partial^2\psi-{g_1}(\psi ^+\psi)\psi /3
  +g_2m\psi] \nonumber\\
  & +  i\lambda g_3 \psi[g_2\psi  ^+\psi-m+h]+  f_{\psi ^+},  \label{eq}
\end{align}
and by the analogous equation for the complex conjugate field $ \psi^{+} $. 
The fields $\psi$, $\psi^{+} $ represent order parameter  (averages of field  operators of 
Bose particles). The field $m$  is a linear combination of  internal energy and
density \cite{Hohenberg} and related to fluctuations of  temperature of the considered system; its
 evolution is described by
\begin{align}
  \partial _tm + \partial_i(v_im)  = -\lambda u\partial ^2[g_2\psi^+\psi -m+h] +i\lambda
  g_3 m+f_m\, .  \label{eq1}
\end{align}
The field $v$ is the fluctuating velocity field (transverse due to incompressibility) and behaves
according to
\begin{align}
  \partial _t v + \partial_i(v_iv) & =  \nu \Delta v-\psi^+\partial[\partial ^2\psi-\frac
  {g_1}3(\psi ^+\psi)\psi+g_2m\psi] \nonumber \\
  & -  \psi\partial[\partial ^2\psi^+-\frac {g_1}3(\psi^+\psi)\psi^++g_2m\psi^+]
  -  m\partial[g_2\psi^+\psi-m+h]+f_v.  \label{eq2}
\end{align}
The random forces  $f_i,i\in\{\psi^+,\psi,m,v\}$ are assumed to be Gaussian random
variables with zero means and correlators $D_i$:
\begin{align}
  \label{eq3}
   D_{\psi}(p, t, t')=\lambda \delta(t-t'),\quad
  D_{m}(p, t,t')=\lambda u p^2 \delta(t-t'), \quad
    D_{v}(p, t, t')=g_4\nu ^3p^{\epsilon -\delta} \delta(t-t').
\end{align}
To analyze the
 model, we use dimensional regularization (see below)
 around its critical dimension four with the standard
 $\varepsilon$-expansion (where $\varepsilon$ is defined by $d=4-\varepsilon$).
 The parameter  $\delta$  measures the deviation from the Kolmogorov regime, i.e. 
the value $\delta =-3$ (and $\varepsilon =1$)   corresponds to the inclusion of  equilibrium  
fluctuations of velocity, and $\delta =4$  defines the regime of developed 
turbulence \cite{Juha1,Juha2,Ant}. We note that 
 Eq.(\ref{eq2}) is the stochastic Navier-Stokes equation with added terms ensuring the existence of an 
equilibrium statistical limit for the proposed model. An important physical fact is that only 
the noise $D_v$ determines which specific hydrodynamic regime is realized.

Our considerations are based on a modification of model E, not only because it is
  relatively simple but also because it was shown in 
 \cite{Dominicis} that this model corresponds to the stable IR-scaling
  regime in model F \cite{Hohenberg}. The standard Martin-Siggia-Rose formalism (MSR)  
\cite{MSR} for the system  (\ref{eq}) leads to the field-theoretic action of the form  
\begin{align}
  S & =  2\lambda {\psi ^+}'\psi
  '-\lambda u m'\partial ^2m' +v'D_vv'+{\psi
  ^+}'\{-\partial_t\psi -\partial _i(v_i\psi)+ \nonumber \\
  & +  \lambda [\partial^2\psi-{g_1}(\psi^+\psi)\psi/3 ] +i\lambda
  g_3 \psi[-m+h] \}+ \nonumber\\
  & +  \psi ' \{-\partial_t \psi^+
  -\partial _i(v_i\psi^+ ) +\lambda [
  \partial^2\psi^+-{g_1}(\psi ^+\psi)\psi^+/3]- \nonumber \\
  & -  i\lambda g_3 \psi ^+[-m+h]  \}+m'\{-\partial _tm-\partial _i(v_im)
  -\lambda u\partial ^2[-m+h]+ \nonumber\\
  & +  i\lambda {g_5}[\psi^+\partial^2\psi-\psi\partial^2\psi^+]\}
   +v'\{-\partial_tv+\nu \Delta v-\partial_i(v_iv)\}   \label{S}\, ,
\end{align}
where integrations over spacetime $(t,{\bm x})$ and summations over repeated
 vector indices are understood.
The terms  
\begin{equation}
  \label{eq4}
  v'\Bigl\{ -\psi^+\partial[\partial
  ^2\psi-\frac {g_1}3(\psi ^+\psi)\psi ] 
  -\psi\partial[\partial ^2\psi^+-\frac {g_1}3(\psi
  ^+\psi)\psi^+ ]- m\partial[ -m+h]\Bigr\}
\end{equation} 
are not included in action (\ref{S}), because it can be shown that they are IR-irrelevant. 

The renormalization of the proposed model was described in detail in \cite{Komarova}. 
 In the renormalization group analysis, the following properties of the model must be applied:
\begin{itemize}
  \item Galilean invariance is present;
  \item nonlocal counterterms of the type  $v'D_vv'$ are absent;
  \item the dimensionless constant  $\nu$ is  expressed in the form $\nu=u_1\lambda$ with
  $u_1$ and is   considered a new charge of the model with its own renormalization constant;
  \item counterterms  of the type $v'\partial _t v$ and $v'(v\partial v)$, are 
  absent, as is usual in developed turbulence;
  \item the derivative in interaction terms  $\phi'\partial_i(v_i\phi)$ can always
   be transferred to the field $\phi '$  or $\phi $ using integration by parts.
\end{itemize}   
In the studied model, the connection with statics is 
violated (because the form of the correlator $D_v$ changes).
Nevertheless, it was shown that the multiplicative renormalization
 can be recovered by adding one new charge at interaction 
 ${g_5}m'(\psi ^+\partial  ^2\psi-\psi \partial ^2 \psi ^+)$, i.e., 
 its bare action is related to the renormalized action 
 by the usual multiplicative relations for the fields and parameters:
\begin{align}
  &  S_R (\varphi) = S(Z_{\varphi} \varphi) , \nonumber \\
  &  Z_{\varphi} \varphi \equiv \left\{ Z_{\psi}\psi, Z_{\psi'}\psi', Z_{\psi^{+}}\psi^{+}, 
  Z_{\psi^{+'}}\psi^{+'}, Z_m m, Z_  {m'}   m', Z_v v, Z_{v'} v' \right \} , \nonumber \\
  &  \lambda_0 = \lambda Z_{\lambda} ,\quad u_0 = u Z_u,\quad u_{10} = u_1 Z_{u_1} ,\quad 
  g_{10} = g_1 \mu^{\varepsilon} Z_{g_1} ,\nonumber \\ 
  &  g_{30} = g_3 \mu^{\frac{\varepsilon}{2}} Z_{g_3} , \quad g_{40} = g_4 \mu^{\delta} Z_{g_4}, 
  \quad { g_{50} = g_5 \mu^  {\frac{\varepsilon}{2}} Z_{g_5}}.
  \label{eq5}
\end{align}
The model is logarithmic for $\varepsilon = \delta = 0$ and the UV
 divergences are manifested in the form of poles in various 
linear combinations of $\varepsilon$ and $\delta$ in dimensional regularization, which
is very convenient for practical calculation\cite{Zinn,Vasiljev2}. These divergences are
 eliminated by introducing the renormalization 
constants. Their explicit form  depends on the choice of the subtraction scheme. 
Of course, universal results are independent of the choice of the particular scheme.
 In the minimal subtraction (MS) scheme only UV divergent terms are subtracted from the Feynman
 diagrams, and we use this scheme in our calculations. The facts indicated above indeed allow proving
 that the renormalized action has the same form as (\ref{S}) and differs by
the renormalized parameters and fields 
 $Z_{\psi +'}$, $Z_{\psi '}$, $Z_{\psi +}$, $Z_{\psi}$,
  $Z_{m'}$, $Z_m$, $Z_{g_1}$, $Z_{g_3}$, ${Z_{g_5}}$, $Z_u$, $Z_{u_1}$, and $Z_{\lambda}$.
 The following relations must be satisfied
 for the renormalization constants of the fields in (\ref{S}):
\begin{align}
   Z_v Z_{v'} = 1, \quad Z_m Z_{m'} = 1,
  \label{eq:renorm_rel}
\end{align}
which are the consequences of the absence of the  renormalization of the
terms $m' \partial m$ and $v' \partial v$.
\section{The UV renormalization}
The RG invariance \cite{Zinn} can be expressed by the differential
 equation $D_{RG} W =0$, where $W$ denotes either the connected
 or the one-particle irreducible (1PI) Green function and the differential part of the
  RG operator is defined as
\begin{align}
  D_{RG} \equiv \mu \frac{\partial}{\partial \mu} \biggl|_{0} = \mu \frac{\partial}{\partial \mu}
   + \sum_{g_i} \beta_{g_i} \frac{\partial}{\partial   g_i} - \sum_a \gamma_a a \frac{\partial}{\partial a}.
   \label{diffoper}
\end{align}
The differentiation is performed at fixed value of the bare parameters, which is indicated by the
 subscript "0". The first summation is over the whole set of charges
  $g_i = \left\{ g_1, g_3, g_4, {g_5}, u, u_1 \right\}$, and
  second is over the set $a=\{\lambda, h\}$. The RG
  functions $\beta_{g_i}$ and $\gamma_F,F = a, g_i$,
  are given by
\begin{align}
  \gamma_F = \mu \frac{\partial\ln Z_F}{\partial \mu} \biggl|_{0},
  \quad \beta_i = \mu \frac{\partial g_i}{\partial \mu} \biggl|_{0}.
  \label{rgfun}
\end{align}

The explicit form of the beta functions follows from this definition and
relations (\ref{eq5}). It is useful to 
 rescale the coupling constants as
\begin{equation}
  g_1/(8\pi ^2) \to g_1,\quad   g_3 /\sqrt{8\pi^2} \to g_3, \quad
  g_4/(8\pi ^2) \to g_4,\quad {g_5 / \sqrt{8\pi^2} \to g_5}.
  \label{res_char}
\end{equation}
 It can be seen from the perturbation expansion that the "real" coupling constants
are the quadratic forms $g_3^2$ and ${g_5^2}$ and not simply $g_3$ and
${g_5}$, whence comes the square root for $g_3$ and ${g_5}$ in (\ref{res_char}).
This fact is also manifested in the fixed-point coordinates because there we expect
that $g_3^2 \varpropto \varepsilon$ and hence $g_3 \varpropto \sqrt{\varepsilon}$ (the same
applies also for the charge ${g_5}$). 
 Using the definitions (\ref{eq5}) and (\ref{diffoper}), we can write $\beta$ functions
  (\ref{rgfun}) in the forms
\begin{align}
   \label{betafun}
  &  \beta _{g_1}=g_1(-\varepsilon -\gamma _{g_1}), &\beta_{g_3}&=g_3(-\varepsilon /2-\gamma _{g_3}), 
  &\beta _{u}&= -u\gamma _{u},\nonumber\\
   &  \beta _{g_4} = g_4( -\delta + 3 \gamma_{\nu}), &\beta_{g_5}& = 
   g_5(-\varepsilon /2-\gamma_{g_5}), &\beta_{u_1}&=-u_1\gamma_{u_1}.
\end{align}
To calculate the renormalization constants in the MS scheme \cite{Zinn}, we must
 the UV-divergent terms (poles in $\epsilon$
and $\delta$ in our case) from the Feynman graph expansion
of the corresponding 1PI functions for the given
 term in action (\ref{S}). We can write 
 these functions schematically in the frequency-momentum representation as
\begin{align}
\label{ren_psi}
\Gamma_{\psi^{+'} \psi^{'}} & =   2 \lambda Z_1 +
  \raisebox{-2.8ex}{ \includegraphics[width=5.2truecm]{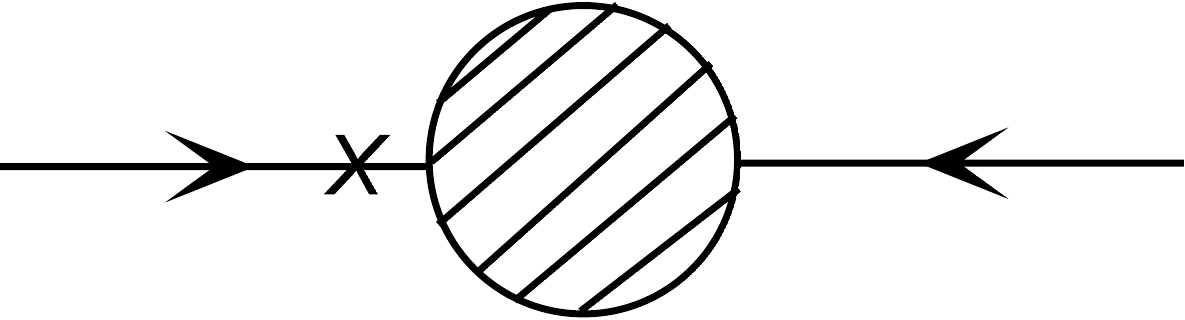}  } \,\,, 
\end{align}  
\begin{align}  
  \Gamma_{m' m'} & =   2 \lambda u p^2 Z_2 +
  \raisebox{-3.2ex}{ \includegraphics[width=5.2truecm]{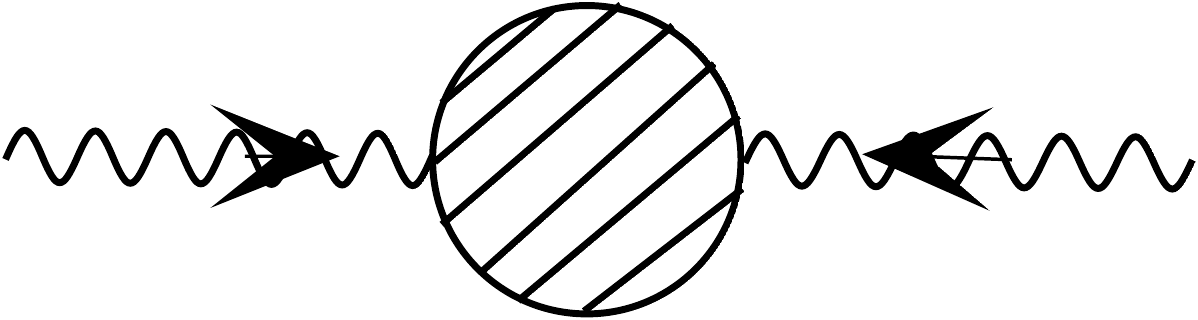}
  } \,\,, 
\end{align}  
\begin{align}  
  \Gamma_{\psi^{+'} \psi} & =   i\omega Z_3 - \lambda p^2 Z_4 +
  \raisebox{-3.0ex}{ \includegraphics[width=5.2truecm]{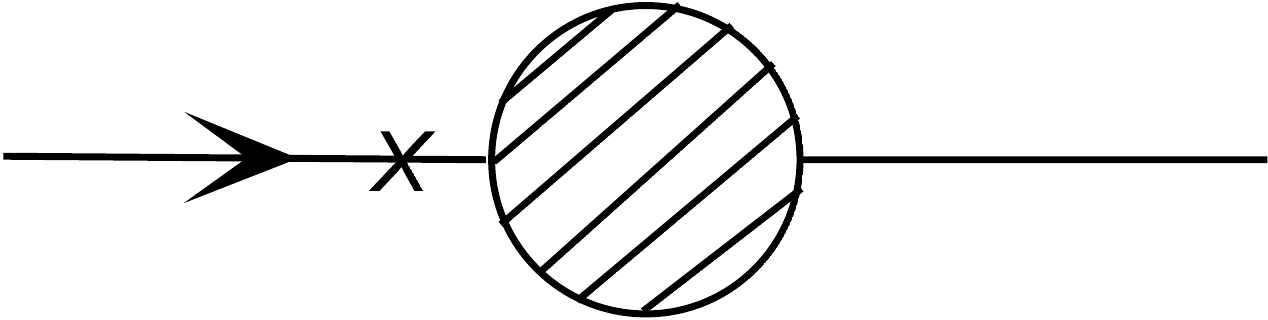}} \,\,,
\end{align}  
\begin{align}  
  \Gamma_{\psi^{+'} \psi^{+} \psi \psi} & =   - \frac{4 \lambda g_1 \mu^{\varepsilon}}{3} Z_5 +
  \raisebox{-4.2ex}{ \includegraphics[width=5.2truecm]{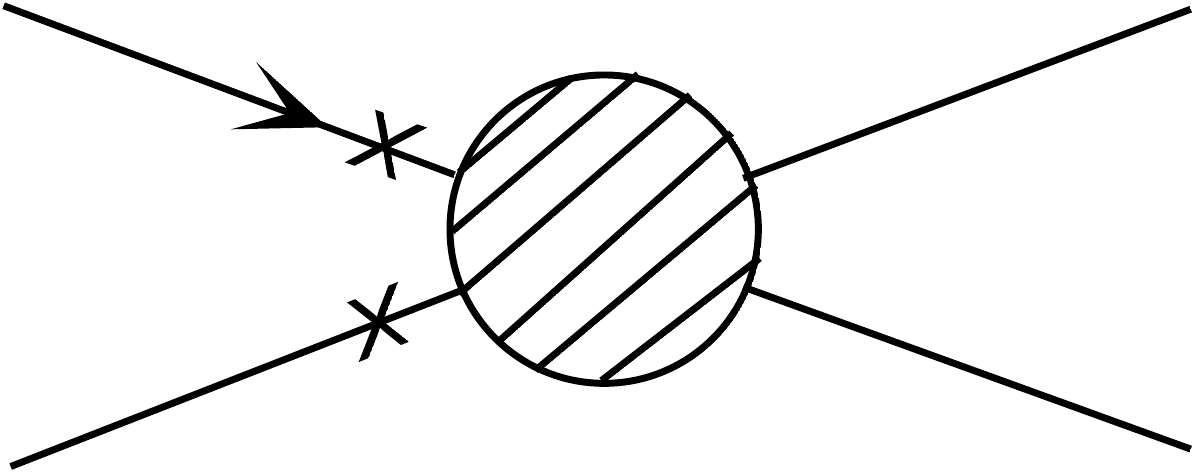}
  } \,\,,
\end{align}  
\begin{align}  
  \Gamma_{\psi^{+'} \psi m} & =   - i \lambda g_3 \mu^{\varepsilon/2} Z_6 +
  \raisebox{-6.2ex}{ \includegraphics[width=5.2truecm]{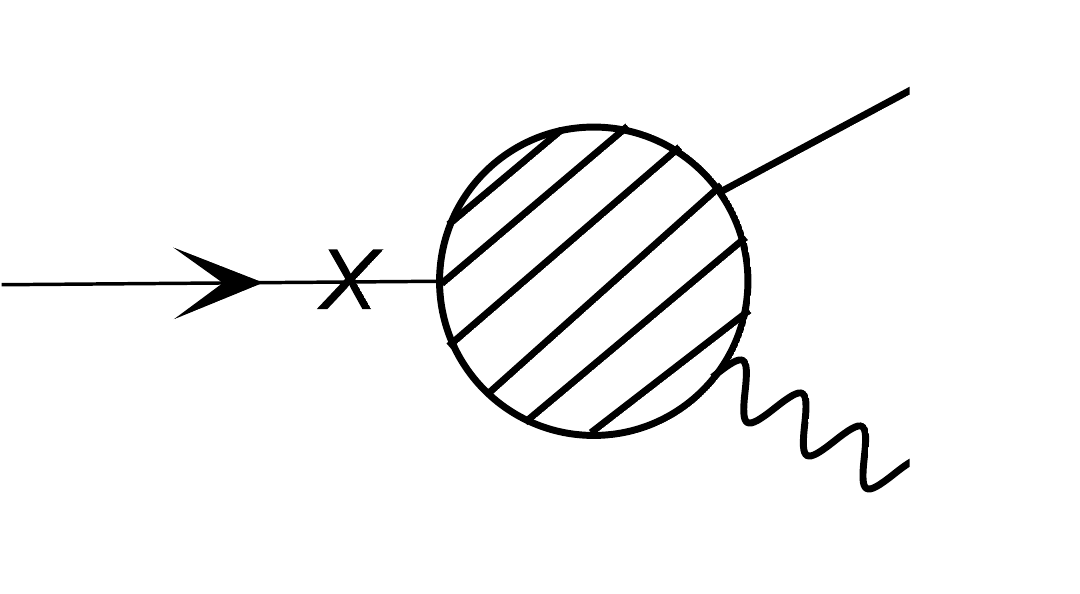}
  } \,\,,
\end{align}  
\begin{align}  
  \Gamma_{m' m} & =   - \lambda u p^2 Z_7 +
  \raisebox{-3.0ex}{ \includegraphics[width=5.2truecm]{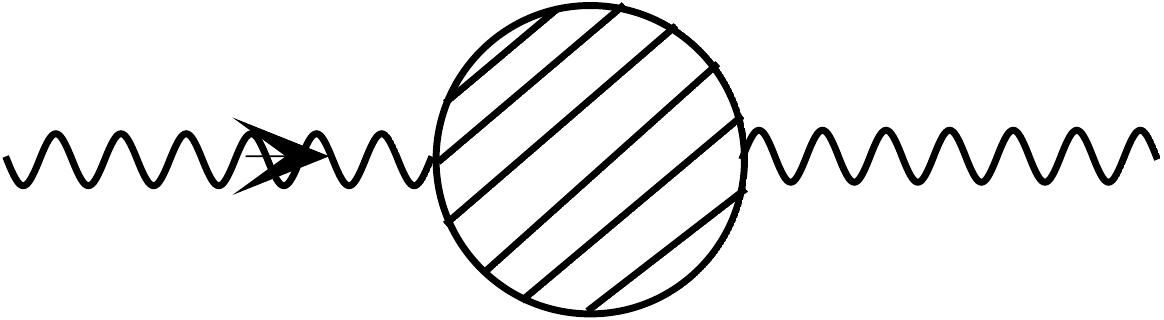}
  } \,\,,
\end{align}  
\begin{align}   
  \Gamma_{m' \psi^{+} \psi} & =   - i \lambda g_5 \mu^{\varepsilon/2} Z_8 +
  \raisebox{-6.2ex}{ \includegraphics[width=5.2truecm]{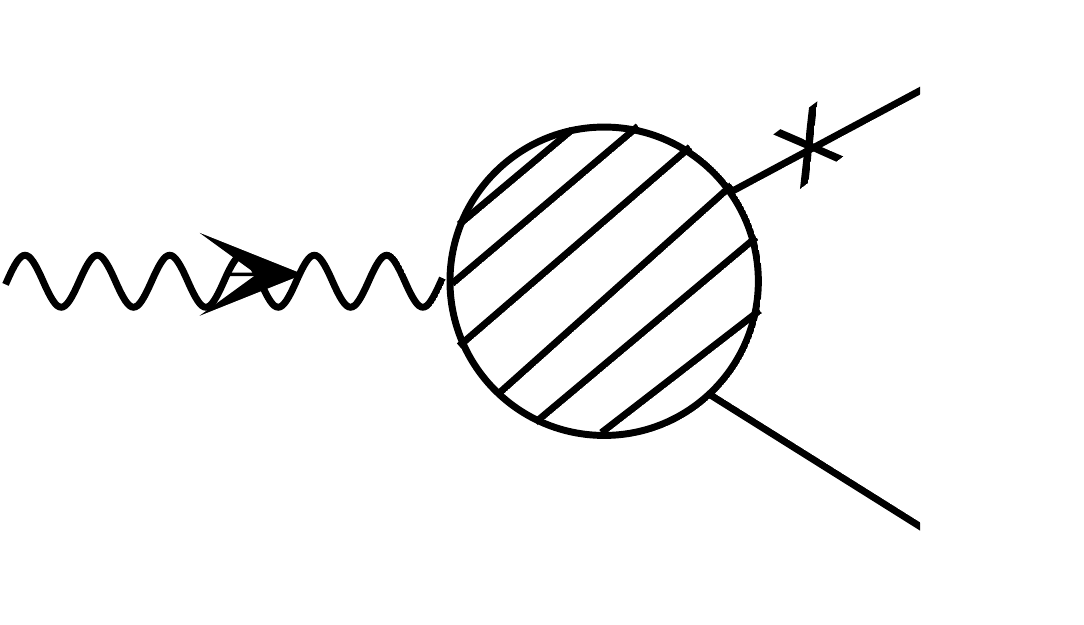}
  },
\end{align}  
\begin{align}    
\label{ren_vv}
   \Gamma_{v' v} & =   - \nu p^2 Z_9 +
  \raisebox{-3.0ex}{ \includegraphics[width=5.2truecm]{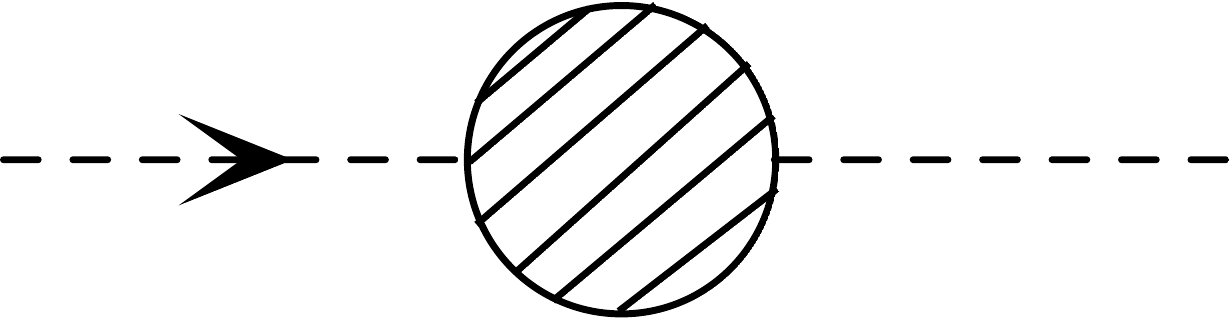}
  } \,\,,
\end{align}
where the solid non-orientable lines denote the legs  formed of $\psi$ fields.
The line with an arrow denotes the response field $\psi'$, the lines with a 
 cross denote the complex-conjugated fields (i.e. $\psi^+$ or ${\psi^{+}}'$), the wavy
 lines denote the fields  $m$ and $m'$, and the dashed lines denote the field $v'$
 (line with arrow) and the field $v$. Shaded blobs represent all possible
 one-loop 1PI Feynman diagrams for the given function.

 Renormalization constants  (\ref{ren_psi})-(\ref{ren_vv})
 are related to the renormalization constants of the parameters
 and fields (\ref{eq5}) via the relations
\begin{align}
  Z_1 & =  Z_{\lambda} Z_{\psi^{+ '}} Z_{\psi '}, 
  &Z_2& = Z_{\lambda} Z_u Z_{m'}^2, &Z_3& = Z_{\psi^{+ '}} Z_{\psi} =
  Z_{\psi^{+ '}} Z_{\psi} Z_v,
  \nonumber \\
  Z_3^{*} & =   Z_{\psi '} Z_{\psi^{+}} = Z_{\psi '} Z_{\psi^{+}} Z_v, 
  &Z_4& = Z_{\psi^{+ '}} Z_{\lambda} Z_{\psi}, 
  &Z_4^{*}& = Z_{\psi'} Z_{\lambda} Z_{\psi^{+}}, 
  \nonumber \\
  Z_5 & =  Z_{\psi^{+ '}} Z_{g_1} Z_{\lambda} Z_{\psi^{+}} Z_{\psi}^2, 
  &Z_5^{*}& = Z_{\psi '} Z_{g_1} Z_{\lambda} Z_{\psi^{+}}^2 Z_{\psi}, 
  &Z_6& = Z_{\psi^{+ '}} Z_{\lambda} Z_{g_3} Z_{\psi} Z_m, 
  \nonumber \\
  Z_6^{*} & =  Z_{\psi '} Z_{\lambda} Z_{g_3} Z_{\psi^{+}} Z_m, 
  &Z_7& = Z_{m'} Z_{\lambda} Z_u Z_m, 
  &Z_8& = Z_{m'} Z_{\lambda} {Z_{g_5}} Z_{\psi^{+}} Z_{\psi},   
  \nonumber \\
  Z_9 & =  Z_{v'} Z_{\nu} Z_v. 
\label{constants}
\end{align}
From these relations, we can be easily obtain
\begin{align}
  Z_{\lambda} & =  Z_4 Z_3^{-1}, &Z_u& = Z_7 Z_3 Z_4^{-1},
  \nonumber \\
  Z_{m'}& = Z_2^{1/2} Z_7^{-1/2}, &Z_{m}& = Z_2^{-1/2} Z_7^{1/2}, 
  \nonumber \\
  Z_{g_3} & =  Z_6 Z_4^{-1} Z_2^{1/2} Z_7^{-1/2},
  &Z_{g_5}& = Z_8 Z_7^{1/2} Z_1 Z_4^{-2} Z_2^{-1/2} Z_3 (Z_3^{*})^{-1},
  \nonumber \\
  Z_{g_1} & = Z_5 Z_1 Z_4^{-2} (Z_3^{*})^{-1}, 
  &Z_{\psi^{+'}}& = Z_{\psi'} = Z_1^{1/2} Z_3^{1/2} Z_4^{-1/2},  \nonumber \\
  Z_{\psi^{+}} & =  Z_{\psi} = Z_1^{-1/2} Z_3^{1/2} Z_4^{1/2}, &Z_{u_1}& = Z_9 Z_3 Z_4^{-1},
  \nonumber \\
  Z_v& = 1, &Z_{\nu}& = Z_9.
\label{constants2}
\end{align}
We can thus obtain the anomalous dimensions $\gamma$ directly from the knowledge of 
renormalization constants $Z_1$-$Z_9$, and in the one-loop
approximation, we obtain the results  
 {
 \begin{align}
    \label{gama1}
   \gamma_{\lambda} & = \frac{3g_4 u_1^2}{8(1+u_1)} + \frac{g_3^2}{(1+u)^3 } + \frac{g_3 g_5 u (2+u)}{(1+u)^3}, \quad \gamma_{g_4} = \frac{g_4}{8 \delta} \nonumber \\
   \gamma_u & =  - \frac{g_3^2}{(1+u)^3} - \frac{g_3 g_5 (u^3 + u^2 - 3u -1)}{2u(1+u)^3} + \frac{3g_4 u_1^2 (1+u_1 - uu_1 - u^2)}{8u(1+u_1)(u+u_1)} \nonumber \\
   \gamma_{g_3} & =  - \frac{3g_4 u_1^2}{8(1+u_1)} - \frac{g_3^2}{(1+u)^3} + \frac{g_5^2}{4u} - \frac{g_3 g_5 (1 + 3u + 11u^2 + 5u^3)}{4u(1+u)^3} \nonumber \\
   \gamma_{g_5} & =  - \frac{3g_4 u_1^2(1+2u+2u_1)}{8(1+u_1)(u+u_1)} - \frac{g_3 g_5 (5u + 23u^2 + 9u^3 - 1)}{4u(1+u)^3} \nonumber \\
                & +  \frac{g_3^2 (2+9u+3u^2)}{2(1+u)^3} - \frac{g_5^2}{4u} \nonumber \\
   \gamma_{g_1} & =  - \frac{3g_4 u_1^2}{4(1+u_1)} - \frac{5 g_1}{3} - \frac{3 g_3^2 g_5 (g_3 - g_5)}{u g_1 (1+u)} + \frac{2 g_3 (1+3u+u^2)(g_3 - g_5)}{(1+u)^3} \nonumber\\ 
   \gamma_{u_1} & =  - \frac{g_3^2}{(1+u)^3} - \frac{g_3 g_5 u (2+u)}{(1+u)^3} + \frac{g_4(1+u_1 - 3u_1^2)}{8 (1+u_1)} \\
   \gamma_{m} & =  \frac{g_3 g_5}{4u} - \frac{g_5^2}{4u}, \quad \gamma_{m'} = - \frac{g_3 g_5}{4u} + \frac{g_5^2}{4u} \nonumber \\
   \gamma_{\psi} & =  \gamma_{\psi^{+}} = \frac{3 g_4 u_1^2}{16 (1+u_1)} - \frac{g_3 (g_3 -g_5)(2+4u+u^2)}{2(1+u)^3 } \nonumber\\
   \gamma_{\psi'} & =  \gamma_{\psi^{+'}} = - \frac{3 g_4 u_1^2}{16 (1+u_1)} + \frac{g_3(g_3-g_5)u(2+u)}{2(1+u)^3 }. \nonumber
 \end{align}}
We note that the limit case $g_3=g_5, g_4=0$ agrees with the results for model E
without velocity fluctuations \cite{Dominicis,Vasiljev2}.
\section{Scaling regimes and fixed points' structure.}
Scaling regimes are associated with fixed points of the corresponding RG functions.
 The fixed points are defined as such points
  $g^{*} = (g_1^{*}, g_3^{*}, g_4^{*}, g_5^{*}, u^{*}, u_1^{*})$ at which all $\beta$ functions 
  vanish simultaneously 
 \begin{align}
    \label{eq6}
    \beta_{g_1}(g^{*})= \beta_{g_3}(g^{*}) = \beta_{g_4}(g^{*})  = \beta_{g_5}(g^{*}) =
    \beta_{u}(g^{*}) = \beta_{u_1}(g^{*}) = 0.
 \end{align}
The type of the fixed point is determined by the eigenvalues of the matrix 
of its first derivatives $\Omega = \left\{ \Omega_{ik} = \partial \beta_i / \partial g_k \right\}$, where
 $\beta_i$ is the full set of $\beta$ functions and $g_k$ is the full
 set of charges $\left\{ g_1, g_3, g_4, g_5, u, u_1 \right\}$. The IR-asymptotic
  behavior is governed by the IR-stable fixed points, for
   which all real parts of eigenvalues of matrix $\Omega$ are positive. Analysis of 
    $\beta$ functions (\ref{betafun}) reveals, that there are several possible regimes
    in the case without thermal fluctuations, i.e., for $g_3=0$. The stable fixed
    points are listed in Table \ref{tab:fp_stab}, and the unstable fixed points
    are listed in Table \ref{tab:fp_unstab}.
\begin{table}[h!]
\centering
\renewcommand{\arraystretch}{1.75}
\begin{tabular}{|c|c|c|c|c|}
  \hline
  FP           & FP1     & FP2         &     FP3 & FP4 \\ \hline
  $g_1$        & $0$     & $0$         &    $\frac{3}{5}\varepsilon$  & $\frac{3}{5}\varepsilon$ \\ \hline
  $g_3$        & $0$     & $0$         &    $\varepsilon^{1/2}$   & $\varepsilon^{1/2}$ \\ \hline
  ${g_5}$ & $0$     & $0$         &    ${\varepsilon^{1/2}}$ 
   & ${\varepsilon^{1/2}}$  \\ \hline
  $g_4 $  & $0$     & $\frac{8\delta}{3}$ & $0$ & $\frac{8\delta}{3}$ \\ \hline
  $u$          & $0$     & $1$         & $1$  & $1$ \\ \hline  
  $u_1$        & $0$     & $\frac{1+\sqrt{13}}{6}$  & $0$ 
  & $0$  \\ \hline
\end{tabular}
  \caption{Stable fixed points}
  \label{tab:fp_stab}
\end{table}

\begin{table}[h!]
\centering
\renewcommand{\arraystretch}{1.75}
\begin{tabular}{|c|c|c|c|c|c|}
  \hline
  FP           & FP5     & FP6         &     FP7 & FP8 & FP9\\ \hline
  $g_1$        & $0$     & $\frac{3\epsilon-2\delta}{5}$         &    $\frac{3\epsilon-2\delta}{5}$  & $0$ & $0$\\ \hline
  $g_3$        & $0$     & $0$         &    $0$   & $\varepsilon^{1/2}$ & $\varepsilon^{1/2}$\\ \hline
  ${g_5}$ & $\frac{\sqrt{2(-19+\sqrt{13})\delta + 18\varepsilon}}{3}$     & $\frac{\sqrt{2(-19+\sqrt{13})\delta + 18\varepsilon}}{3}$         &    $0$ 
   & ${\varepsilon^{1/2}}$ & ${\varepsilon^{1/2}}$ \\ \hline
  $g_4 $  & $\frac{8\delta}{3}$     & $\frac{8\delta}{3}$ & $\frac{8\delta}{3}$ & $0$ & $\frac{8\delta}{3}$\\ \hline
  $u$          & $1$     & $1$         & $1$  & $1$ & $1$\\ \hline  
  $u_1$        & $\frac{1+\sqrt{13}}{6}$     & $\frac{1+\sqrt{13}}{6}$  & $\frac{1+\sqrt{13}}{6}$ 
  & $0$ & $0$ \\ \hline
\end{tabular}
  \caption{Unstable fixed points}
  \label{tab:fp_unstab}
\end{table}
The trivial Gaussian-like fixed point FP1 is IR-stable for $\varepsilon < 0$ and $\delta <0$ and
corresponds
to the model without any nontrivial interactions. The fixed point FP2 is a IR-stable in the region
given by the inequalities $\delta > 0$ and $\delta > \frac{3}{2}\varepsilon$ and corresponds to the 
turbulent regime (because $g_4^{*} \neq 0$,$\delta =4$ and $\gamma_{\nu}^{*} = \frac{\delta}{3}$). 

The fixed points FP3 and FP4 differ only by the value of the charge $g_4^{*}$.
The hydrodynamic fluctuations of the velocity field are IR irrelevant for FP3 and relevant,
 for the FP4. The fixed point FP3 is stable in the region where $\delta < 0,\varepsilon>0$ 
 and FP4 is stable for $\delta > 0,\delta < \frac{3}{2}\varepsilon$. 
 Comparing FP8 and FP9 with their analogues FP3 and FP4,
 we can see that the absence of the interaction term
   $\psi^{+'} \psi^{+} \psi^2$ leads to system instability.
   We expect that this behavior can be explained by the disordering effect
   due to thermal fluctuations (charge $g_3\neq 0$) because there
   are no other  interactions between the relevant degrees of 
   freedom (fields of the type $\psi$) that could stabilize system.
   
Briefly examining the common properties of the fixed points FP5-FP7,
 we see that regardless of the presence of the interaction $\psi^{+'} \psi^{+} \psi^2$,  
 velocity fluctuations destabilize IR behavior. 

{The charges $u$ and $u_1$ do not play the role of expansion parameters and it 
 therefore seems reasonable to consider specific limits as their values tend
to infinity.
 We consider the case where $u \to \infty$ (case I) in Table \ref{tab:fpu}. To analyze this regime,
 we introduce new variables
$w \equiv 1/u$, $f_3 \equiv g_3^2/u$, and $f_5 \equiv g_5^2/u$.
Their beta functions have the form $\beta_w = w \gamma_u$, $\beta_{f_3} = f_3[-\varepsilon
+ \gamma_u - 2\gamma_{g_3}]$ and
$\beta_{f_5} = f_5[-\varepsilon + \gamma_u - 2\gamma_{g_5}]$. The fixed points FP1$^{I}$ is
 Gaussian (free).
 The fixed points  FP2$^{I}$ and FP3$^{I}$ differ only by the value of
$g_4^{*}$. The fixed point $FP4^{I}$ corresponds to the turbulent regime where the
interaction
$\psi^{+'} \psi^{+} \psi^2$ is relevant.
 The last fixed point $FP5^{I}$ is case without thermal fluctuations ($f_3 = 0$).

 We consider another limit case where $u_1 \to \infty$ (case II) in
 Table \ref{tab:fpu1}. In this case, we introduce new variables $w_1 = 1/u_1$ and $f_4 = g_4 u_1$.
 The corresponding beta functions have the  forms $\beta_{w_1} = w_1 \gamma_{u_1}$ and
$\beta_{f_4} = f_4 [-\delta + 3 \gamma_{\nu} - \gamma_{u_1}]$. From Table \ref{tab:fpu1}, we
 again see that the only difference between FP2$^{II}$ and FP3$^{II}$ is the charge $g_4^{*}$, 
 and FP4$^{II}$ corresponds to a turbulent regime. The fixed point FP5$^{II}$ corresponds to a
 nontrivial IR-scaling regime without thermal fluctuations.

Finally, we analyze the case where both charges $u$ and $u_1$ tend to infinity simultaneously
  (see Table \ref{tab:fpuu1}). In the FP2$^{III}$ regime, the presence of the interaction 
term $\psi^{+'} \psi^{+} \psi^2$ is irrelevant, unlike for the FP3$^{III}$.The fixed point
 FP4$^{III}$ corresponds to the turbulent regime with the interaction
$\psi^{+'} \psi^{+} \psi^2$, while that interaction is irrelevant in the regime FP5$^{III}$.}
\begin{table}[ht!]
\centering
\renewcommand{\arraystretch}{1.75}
\begin{tabular}{|c|c|c|c|c|c|}
  \hline
  FP           & FP1$^{I}$     & FP2$^{I}$  &     FP3$^{I}$ & FP4$^{I}$ & FP5$^{I}$\\ \hline
  $g_1$        & $0$     & $\frac{3\epsilon}{5}$         &    $\frac{3\epsilon}{5}$  & $\frac{1}{5}(3\epsilon-2\delta)$ & $\frac{1}{5}(3\epsilon-2\delta)$\\ \hline
  $f_3$        & $0$     & $\frac{2\epsilon}{3}$         &    $\frac{2\epsilon}{3}$   & $0$ & $0$\\ \hline
  $f_5$        & $0$     & $\frac{2\epsilon}{3}$         &    $\frac{2\epsilon}{3}$ & $0$ & $2\epsilon - 2\delta$ \\ \hline
  $g_4$        & $0$     & $0$ & $\frac{8\delta}{3}$ & $\frac{8\delta}{3}$ & $\frac{8\delta}{3}$\\ \hline
  $w$          & $0$     & $0$         & $0$  & $0$ & $0$\\ \hline  
  $u_1$        & $0$     & $0$  & $0$ 
  & $\frac{1}{6}(1+\sqrt{13})$ & $\frac{1}{6}(1+\sqrt{13})$ \\ \hline
\end{tabular}
  \caption{Fixed points for limiting case $u \to \infty$}
  \label{tab:fpu}
\end{table}

\begin{table}[ht!]
\centering
\renewcommand{\arraystretch}{1.75}
\begin{tabular}{|c|c|c|c|c|c|}
  \hline
  FP           & FP1$^{II}$     & FP2$^{II}$  &     FP3$^{II}$ & FP4$^{II}$ & FP5$^{II}$\\ \hline
  $g_1$        & $0$     & $0$         &    $\frac{3\epsilon}{5}$  & $\frac{3}{5}(\epsilon-2\delta)$ & $\frac{3}{5}(\epsilon-2\delta)$\\ \hline
  $g_3$        & $0$     & $0$         &    $0$   & $0$ & $0$\\ \hline
  $g_5$        & $0$     & $\sqrt{2\epsilon}$         &    $\sqrt{2\epsilon}$ & $0$ & $\sqrt{2(\epsilon-4\delta)}$ \\ \hline
  $f_4$        & $0$     & $0$ & $0$ & $\frac{8\delta}{3}$ & $\frac{8\delta}{3}$\\ \hline
  $u$          & $0$     & $1$         & $1$  & $1$ & $1$\\ \hline  
  $w_1$        & $0$     & $0$  & $0$ & $0$ & $0$ \\ \hline
\end{tabular}
  \caption{Fixed points for limiting case $u_1 \to \infty$}
  \label{tab:fpu1}
\end{table}

\begin{table}[ht!]
\centering
\renewcommand{\arraystretch}{1.75}
\begin{tabular}{|c|c|c|c|c|c|}
  \hline
  FP           & FP1$^{III}$     & FP2$^{III}$  &     FP3$^{III}$ & FP4$^{III}$ & FP5$^{III}$\\ \hline
  $g_1$        & $0$     & $0$         &    $\frac{3\epsilon}{5}$  & $\frac{3}{5}(\epsilon-2\delta)$ & $\frac{3}{5}(\epsilon-2\delta)$\\ \hline
  $f_3$        & $0$     & $\frac{2 \epsilon}{3}$         &    $\frac{2 \epsilon}{3}$   & $0$ & $0$\\ \hline
  $f_5$        & $0$     & $\frac{2 \epsilon}{3}$         &    $\frac{2 \epsilon}{3}$ & $0$ & $2(\epsilon-3\delta)$ \\ \hline
  $f_4$        & $0$     & $0$ & $0$ & $\frac{8\delta}{3}$ & $\frac{8\delta}{3}$\\ \hline
  $w$          & $0$     & $0$         & $0$  & $0$ & $0$\\ \hline  
  $w_1$        & $0$     & $0$  & $0$ & $0$ & $0$ \\ \hline
\end{tabular}
  \caption{Fixed points for limiting case case $u \to \infty$and $u_1 \to \infty$}
  \label{tab:fpuu1}
\end{table}
The last most, nontrivial case corresponds to the situation  where
 all charges have non-zero values. But because the
 structure of the $\gamma$-functions is cumbersome, we have not yet found the coordinates
 of this fixed point and its region of stability. Of course, from other
 fixed points, we know where to expect such a stability region. In the near
 future, we hope to confirm our expectations by direct
 numerical calculations.
\section{Conclusion}
We have studied model E was studied in the vicinity of the critical point of the phase transition from 
the normal to the superfluid phase with both critical and velocity fluctuations taken into account.
 We showed that the model can be made multiplicatively renormalizable by adding a new charge in the
 interaction part of the action. We calculated the renormalization constants and
 RG functions up to the first
 order (one-loop) in the perturbation theory and partly analyzed the fixed-point structure. 
 Our main observation is that incorporation of velocity fluctuations destabilizes
 the critical behavior.\\

{The work was supported by grant RFBR No.12-02-00874-a, by VEGA grant
1/0222/13 of the Ministry of Education, Science, Research and Sport of the Slovak Republic, by Center
of Excellency for Nanofluid of IEP SAS and by National Scholarship Program
of the Slovak Republic.} This article was also created by implementation
of the Cooperative phenomena and phase transitions in nanosystems with
perspective utilization in nano-and biotechnology project No 26220120033, No 26110230061
and {No 26220120021}.
Funding for the operational research and development program was provided by the
European Regional Development Fund.

     \end{document}